\documentstyle[aps,epsf,preprint]{revtex}
\tightenlines
\begin{document}
\draft
\title{Radiative decays of \boldmath{$\phi$}-meson about
nature of light scalar resonances}
\author{N.N. Achasov
\thanks{Email address: achasov@math.nsc.ru}}
\address{Laboratory of Theoretical Physics,
 Sobolev Institute for Mathematics,
  Novosibirsk, 630090, Russia}
  \date{\today}
 \maketitle
\begin{abstract}
We show on gauge invariance grounds that the fine threshold
phenomenon is discovered in the radiative decays $\phi\to\gamma
a_0\to\gamma\pi^0\eta$ and
 $\phi\to\gamma f_0\to\gamma\pi^0\pi^0$. This enables to conclude
 that production of the lightest scalar mesons $a_0(980)$ and
 $f_0(980)$ in these decays
  is caused by the four-quark transitions, resulting in strong restrictions on the
 large $N_C$ expansions of the decay amplitudes. The analysis
 shows that these constraints give new evidences in favor
 of the four-quark nature of $a_0(980)$ and $f_0(980)$ mesons.
\end{abstract}
\vspace*{1cm}

 \section{Introduction}

  The lightest scalar mesons
$a_0(980)$ and $f_0(980)$, discovered more than thirty years ago,
became the hard problem for the naive quark-antiquark ($q\bar q$)
model from the outset. Really, on the one hand the almost exact
degeneration of the masses of the isovector $a_0(980)$ and
isoscalar $f_0(980)$ states revealed seemingly the structure
similar to the structure of the vector $\rho$ and $\omega$ mesons
, and on the other hand the strong coupling of $f_0(980)$ with the
$K\bar K$ channel pointed unambiguously to a considerable part of
the  strange quark pair $s\bar s$ in the wave function of
$f_0(980)$.

In 1977 R.L. Jaffe  noted that in the MIT bag model, which
incorporates confinement phenomenologically, there are light
four-quark scalar states \cite{jaffe}. He suggested also that
$a_0(980)$  and $f_0(980)$   might be these states with symbolic
structures $a^0_0(980)=(us\bar u\bar s - ds\bar d\bar s)/\sqrt{2}$
and $f_0(980)=(us\bar u\bar s + ds\bar d\bar s)/\sqrt{2}$. From
that time $a_0(980)$ and $f_0(980)$ resonances came into beloved
children of the light quark spectroscopy, see, for example,
\cite{montanet,achasov-84,achasov}.

Ten years later we showed \cite{achasov-89} that the study of the
radiative decays $\phi\to\gamma a_0\to\gamma\pi\eta$ and
$\phi\to\gamma f_0\to \gamma\pi\pi$ can shed light on the problem
of  $a_0(980)$ and $f_0(980)$ mesons. Over the next ten years
before experiments (1998) the question was considered from
different points of view
\cite{bramon,achasov-97,achasov-97a,achasov-97b,achasov-98}.

Now these decays have been studied not only theoretically but also
experimentally.

The first measurements have been reported by the SND
\cite{snd-fit,snd-ivan} and CMD-2 \cite{cmd} Collaborations which
obtain the following branching ratios
$$BR(\phi\to\gamma\pi^0\eta)=
(0.88\pm 0.14\pm 0.09)\cdot10^{-4}\
\mbox{\cite{snd-fit}\,(2000)},$$
$$ BR(\phi\to\gamma\pi^0\pi^0)= (1.221\pm 0.098\pm 0.061)\cdot10^{-4}\
\mbox{\cite{snd-ivan}\,(2000)},$$
 $$BR(\phi\to\gamma\pi^0\eta)=(0.9\pm 0.24\pm 0.1)\cdot10^{-4}\ \mbox{\cite{cmd}},$$
$$BR(\phi\to\gamma\pi^0\pi^0)=(0.92\pm0.08\pm0.06)\cdot10^{-4}\
\mbox{\cite{cmd}}.$$
 More recently the KLOE Collaboration has measured
\cite{kloea0,kloef0}
$$ BR(\phi\to\gamma\pi^0\eta)=(0.851\pm0.051\pm0.057)\cdot 10^{-4}\
\mbox{in}\  \eta\to\gamma\gamma\ \mbox{\cite{kloea0}},$$
$$BR(\phi\to\gamma\pi^0\eta)=(0.796\pm0.060\pm0.040)\cdot10^{-4}\
\mbox{in}\ \eta\to\pi^+\pi^-\pi^0\ \mbox{\cite{kloea0}},$$
$$BR(\phi\to\gamma\pi^0\pi^0)= (1.09\pm0.03\pm0.05)\cdot10^{-4}\
\mbox{\cite{kloef0}},$$ in agreement with the Novosibirsk data
\cite{snd-fit,snd-ivan,cmd} but with a considerably smaller error.

Note that the isovector $a_0(980)$ meson is produced in the
radiative $\phi$ meson decay
 as intensively as the well-studied $\eta '(958)$ meson containing $\approx 66\% $ of $s\bar s$,
  responsible for the decay ($\phi\approx s\bar s\to\gamma s\bar s\to\gamma \eta '(958)$).
 It is a clear qualitative argument
for the presence of the $s\bar s$ pair in the isovector $a_0(980)$
state, i.e., for its four-quark nature.

\begin{figure}
\centerline{ \epsfxsize=14cm \epsfysize=4cm \epsfbox{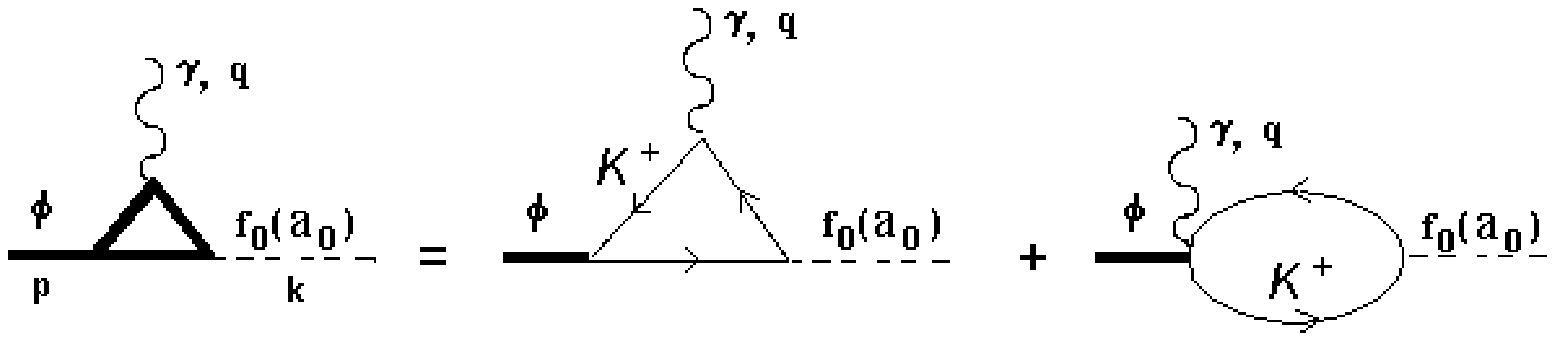}}
 \caption{Diagrams of the $K^+K^-$ loop model.  }
\label{model}
\end{figure}

Since the one-loop model  $\phi\to K^+K^-\to\gamma a_0$ and
$\phi\to K^+K^-\to\gamma f_0$, see Fig. \ref{model},  suggested at
basing the experimental investigations \cite{achasov-89}, is used
in the data treatment \cite{snd-fit,snd-ivan,cmd,kloea0,kloef0},
the question on the mechanism of the scalar meson production in
the $\phi$ radiative decays is put into the shade.

We show below in Section II that the present data give the
conclusive arguments in favor of the $K^+K^-$ loop transition as
the principal mechanism  of $a_0(980)$ and $f_0(980)$ meson
production in the $\phi$ radiative decays.

 In Section III we show that the knowledge of this mechanism
allows to conclude that the production of $a_0(980)$ and
$f_0(980)$ in the $\phi$ radiative decays is caused by the
four-quark transitions. This constrains the large
 $N_C$ expansions of the decay amplitudes and gives new impressive evidences in favor
 of the four-quark nature of $a_0(980)$ and $f_0(980)$
  \cite{jaffe,achasov-84,achasov,achasov-89,achasov-97,black,josef,achasov-01}.

Section IV contains concluding remarks.

\section{The mechanism of the \lowercase{\boldmath{$a_0(980)$}} and
\lowercase{\boldmath{$f_0(980)$}} production in the
\lowercase{\boldmath{$\phi$}} radiative decays}

\begin{figure}
 \centerline{\epsfxsize=14cm \epsfysize=9cm
\epsfbox{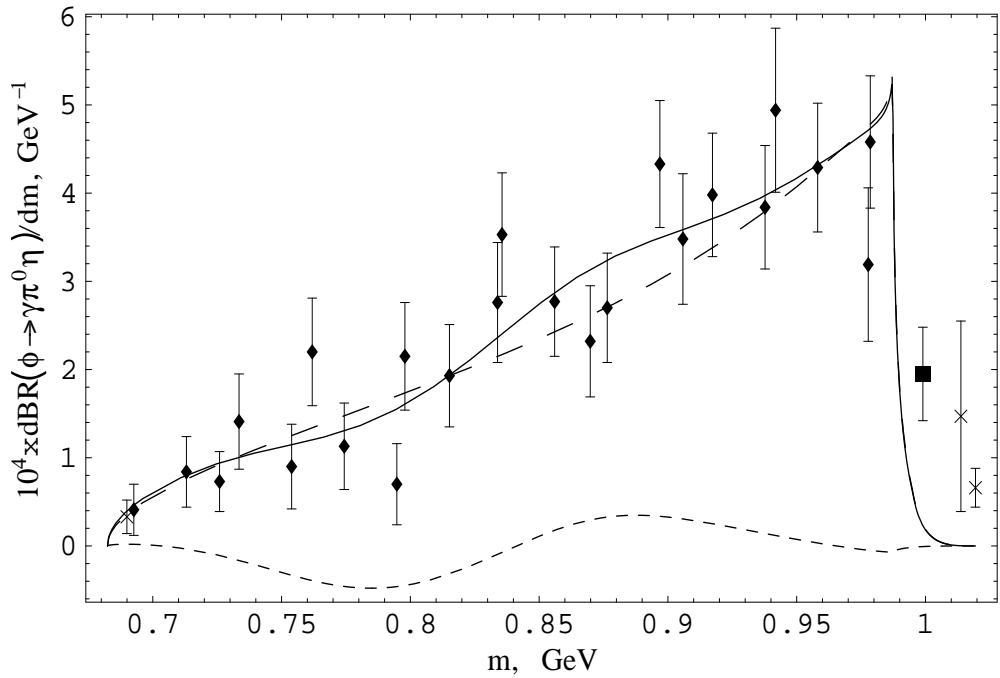}} \caption{Fitting of $10^{4}\times
dBR(\phi\to\gamma\pi^0\eta)/dm$ with the background  is shown with
the solid line. The signal contribution and  interference
contributions are shown with the dashed and dotted lines
respectively. The data are from the KLOE detector.}
\label{myfiga0}
\end{figure}

\begin{figure}
 \centerline{\epsfxsize=14cm \epsfysize=9cm \epsfbox{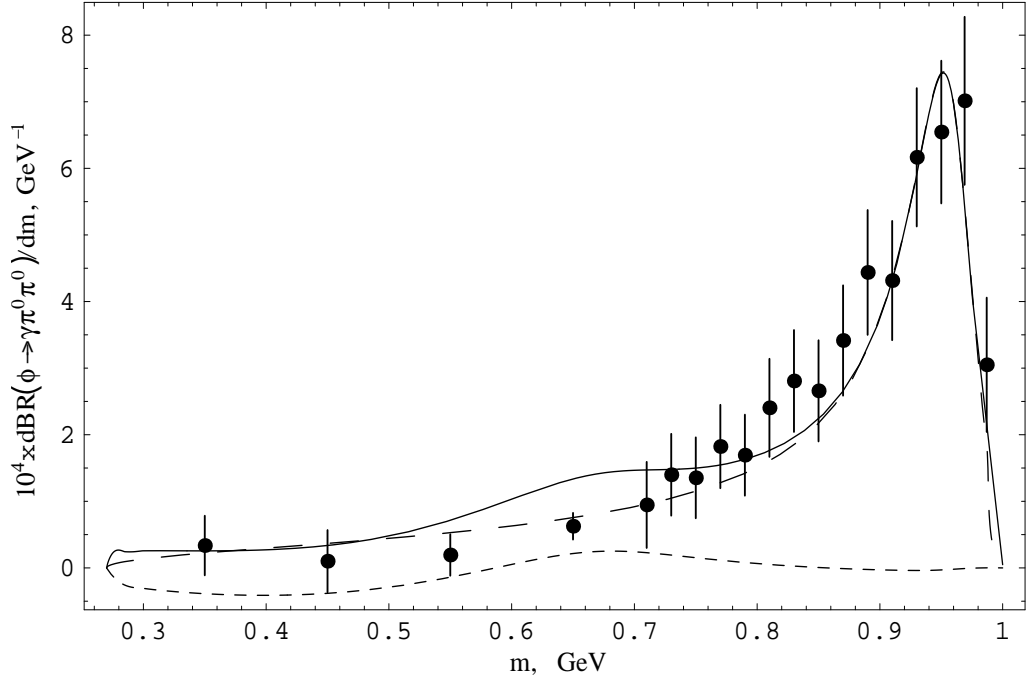}}
 \caption{ Fitting of $ 10^{4}\times dBR(\phi\to\gamma\pi^0\pi^0)/dm$ with
the background is shown with the solid line, the signal
contribution is shown with the dashed line. The dotted line is the
interference term. The data are from the SND detector.
  } \label{figf0}
\end{figure}

In Figs. \ref{myfiga0} and \ref{figf0} are shown the KLOE data on
$\phi\to\gamma\pi^0\eta$ \cite{kloea0} and the SND data on
$\phi\to\gamma\pi^0\pi^0$ \cite{snd-ivan} (2000) respectively. The
 excitations of the $a_0(980)$ resonance, similar to the one in Fig.
 \ref{myfiga0}, are observed also
by the SND \cite{snd-fit} and CMD-2 \cite{cmd} Collaboraions. The
 excitations of the $f_0(980)$ resonance, similar to the one in Fig.
 \ref{figf0}, are observed also
by the CMD-2 \cite{cmd} and KLOE \cite{kloef0} Collaborations.
 The data are described in the model
$\phi\to(\gamma a_0+\pi^0\rho)\to\gamma\pi^0\eta$ and
$\phi\to(\gamma f_0+\pi^0\rho)\to\gamma\pi^0\pi^0$, see details in
Ref. \cite{achasov-01,kiselev}.

As Figs. \ref{myfiga0}  and \ref{figf0}
 \footnote{A.V. Kiselev noted kindly that in Ref. \cite{achasov-01} the solid
curve at $m < 780$ MeV is drew incorrectly. He also kindly
prepared the correct figure.}
 suggest, the $\phi\to\gamma
a_0\to\gamma\pi^0\eta$ process dominates everywhere over the
region of the $\pi^0\eta$ invariant mass $m_{\pi^0\eta}= m$ and
the $\phi\to\gamma f_0\to\gamma\pi^0\pi^0$ process dominates in
the resonance region of the $\pi^0\pi^0$ system, the $\pi^0\pi^0$
invariant mass $m_{\pi^0\pi^0}=m > 780$ MeV.

  Notice that the OZI suppressed $BR(\phi\to\rho\pi)=0.14$ is caused by the
large phase space (including the $P$-wave factor) of the $\rho\pi$
states in comparison with the one of the $K\bar K$ states in the
$\phi\to K\bar K$ decay. As for the coupling constants,
 $\left |g_{\phi\rho\pi}/g_{\phi K\bar K}\right |^2\approx 0.03$
GeV$^{-2}$ (please take into account also that a typical energy
scale $<$ 1 GeV in our case). As a result at tree level
approximation
$BR(\phi\to\rho\pi\to\gamma\pi^0\eta)/BR(\phi\to\gamma\pi^0\eta)\approx
0.07$ and
$BR(\phi\to\rho\pi\to\gamma\pi^0\pi^0)/BR(\phi\to\gamma\pi^0\pi^0)\approx
0.15$, see \cite{achasov-01,kiselev}, at a sacrifice of the large
phase spaces (including the $\omega^3=(\mbox{photon energy})^3$
factor from gauge invariance) in comparison with the ones in the
$\phi\to\gamma a_0\to\gamma\pi^0\eta$ and $\phi\to\gamma
f_0\to\gamma\pi^0\pi^0$ decays. Taking into account rescatterings
$\phi\to\rho\pi\to\gamma(\pi^0\eta\to\pi^0\eta)$ and
$\phi\to\rho\pi\to\gamma(\pi\pi\to\pi^0\pi^0)$ does not change
situation considerably, see Ref. \cite{oset-03}.

\begin{figure}
\centerline{\epsfxsize=14cm \epsfysize=8.5cm \epsfbox{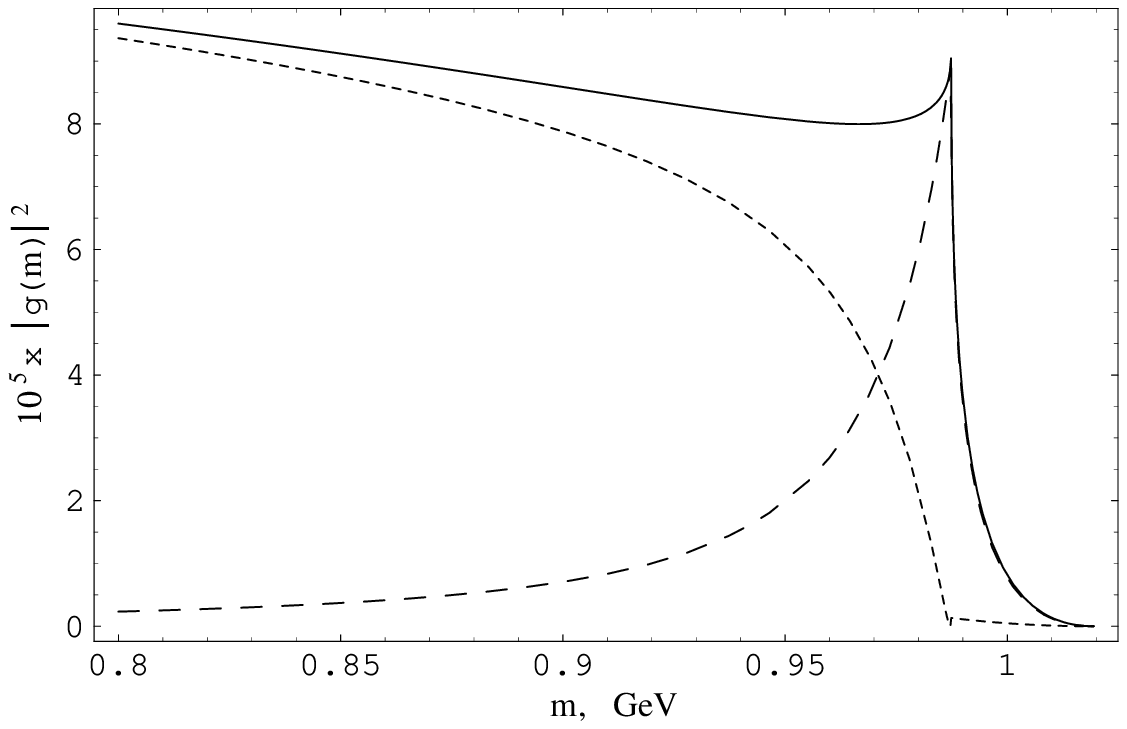}}
 \caption{ The function $|g(m)|^2$ is drawn with the solid line. The contribution of the
  imaginary part is drawn with the dashed line. The contribution of the real part
   is drawn with the dotted line.} \label{g}
\end{figure}

 Notice also that the conclusion of Ref. \cite{mrp-03}, that in the $\phi\to\gamma\pi^0\pi^0$ the
$K\bar K$ intermediate states  dominate for 950-980 MeV, whereas
the $\pi\pi$ intermediate states dominate for 700-900 MeV, is
based on a mistaken hypothesis. The basic equation of the
investigation in Ref. \cite{mrp-03}, Eq. (3), with the real
$\alpha_1(s)$ and $\alpha_2(s)$ (as the authors supposed) is not
correct. The point is that in the case under consideration there
are discontinuities not only in the $\pi\pi$ ($\pi\pi$ and $K\bar
K$ intermediate states)  channel but also in the $\phi$
(intermediate $K\bar K$ and $\rho\pi$ states) and $\gamma\pi$
(intermediate $\rho$ resonance) channels. Let us consider this
question in detail. As is well know the imaginary part of a
amplitude is a half of the sum of all possible discontinuities in
all channels. According to Ref. \cite{mrp-03}, the discontinuities
of the $\phi\to\gamma\pi^0\pi^0$ amplitude are caused only by the
$\pi\pi\to\pi\pi$ and $\pi\pi\to K\bar K$ amplitudes, that is,
only by the $\pi\pi$ intermediate states at the invariant mass of
the $\pi\pi$ state less than $2M_K$, $m=m_{\pi\pi}\leq 2M_K$. But
it is not the case. The two step real intermediate physical
process $\phi\to K^+K^-$ and then $K^+K^-\to\gamma\pi^0\pi^0$
causes the discontinuity in the $\phi$ channel at any
$m=m_{\pi\pi}$. This contribution is crucial, as one can see from
the behaviour of the contribution of the imaginary part in  Fig.
\ref{g}. Emphasize that the above the two step real physical
process is not foreseen in Ref. \cite{mrp-03} at all. At
$m=m_{\pi\pi}\geq 2M_K$ the new two step real intermediate
physical process $\phi\to\gamma K^+K^-$ and then
$K^+K^-\to\pi^0\pi^0$ opens and causes the discontinuity, which
compensates the previous one, so that the sum of these two
discontinuities vanishes at $\omega=0$ (or $m=m_{\pi\pi}=m_\phi$)
in agreement with gauge invariance, as one can see from the
behaviour of the contribution of the imaginary part in  Fig.
\ref{g}. Certainly, the discontinuity in the $\phi$ channel,
caused by the real $\rho\pi$ intermediate states, is not
essential, as discussed above. The $\rho$ contribution in the
$\gamma\pi$ channel ($\phi\to\rho^0\pi^0\to\gamma\pi^0\pi^0$) is
discussed above also. Its value in the branching ratio is
calculated to within 20\% \cite{achasov-01,kiselev} because both
the $\phi\to\rho\pi$ decay and the $\rho\to\gamma\pi$ one are
studied well enough. The absence of this contribution in the KLOE
data is caused by  the experimental cutoff, which removes the
$\omega\pi^0\to\gamma\pi^0\pi^0$ events. The point is that  the
$\rho^0\pi^0\to\gamma\pi^0\pi^0$ events are in the same region of
the $\pi\pi$ mass ($m=m_{\pi\pi}< 630$ MeV) as the
$\omega\pi^0\to\gamma\pi^0\pi^0$ ones but they are only a small
fraction of all $\gamma\pi^0\pi^0$ events (less than 10\%). As a
consequence the most part of them is removed by the above cutoff.

 As for a contact $\phi\to\gamma K^0 \bar K^0$
interaction, which was suggested in Ref. \cite{oller-03} to
improve fitting data in a specific model, this could only replace
"the dominance of the $K^+K^-$ intermediate states" by
 "the dominance of the $K\bar K$ intermediate states".
 Notice, however, that the calculation
 of the $\phi\to\gamma K^0 \bar K^0$
interaction through the $\phi\to K^\ast \bar K\to\gamma K\bar K$
loops in a chiral theory requires the introduction of a conterterm
which can be fixed now only by experiment. As for the experiment
data, they can be described also without
 a $\phi\to\gamma K^0 \bar K^0$  contribution,
 see, Refs. \cite{achasov-01,kiselev} and Figs. \ref{myfiga0} and \ref{figf0}.

The resonance mass spectrum is of the form
\footnote{Notice that
in Ref. \cite{achasov-01} we took into account the mixing of
$f_0(980)$ meson with another scalar isoscalar resonance, see also
Ref. \cite{achasov-97}, but such a complication is not essential
for the present investigation.}

\begin{eqnarray}
&&S_R(m)=d\Gamma(\phi\to\gamma R\to\gamma ab\,,\, m)/dm\nonumber\\
&&=\frac{2}{\pi}\frac{m^2\Gamma(\phi\to\gamma R\,,\,m)\Gamma(R\to
ab\,,\,m)}{|D_R(m)|^2} = \frac{4|g_R(m)|^2\omega (m)
p_{ab}(m)}{3(4\pi)^3m_{\phi}^2}\left |\frac{g_{Rab}}{D_R(m)}\right
|^2,
 \label{spectrumR}
\end{eqnarray}
where $R = a_0\ \mbox{or}\ f_0$ and $ab=\pi^0\eta\ \mbox{or}\
\pi^0\pi^0$ respectively, $\omega (m)=(m_{\phi}^2-m^2)/2m_{\phi}$
is the photon energy in the $\phi$ meson rest frame, $p_{ab}(m)$
is the modulus of the $a$ or $b$ particle momentum in the $a$ and
$b$ mass center frame, $g_{Rab}$ is the coupling constant,
$g_{f_0\pi^0\pi^0}=g_{f_0\pi^+\pi^-}/\sqrt{2}$, $D_R(m)$ is the
$R$ resonance propagator the form of which everywhere over the $m$
region can be find in \cite{achasov-89,achasov-80,achasov-01a},
$g_R(m)$ is the invariant amplitude that describes the vertex of
the $\phi (p)\to\gamma (k) R(q)$ transition with $q^2=m^2$. This
is precisely the function which is the subject of our
investigation.

By  gauge invariance, the transition amplitude is proportional to
the electromagnetic field strength tensor $F_{\mu\nu}$ (in our
case to the electric field in the $\phi$ meson rest frame):
\begin{eqnarray}
&& A\left [\phi(p)\to\gamma (k) R(q)\right ]= G_R(m)\left [p_\mu
e_\nu(\phi) - p_\nu e_\mu(\phi)\right]\left [k_\mu e_\nu(\gamma) -
k_\nu e_\mu(\gamma)\right],
 \label{gauge}
\end{eqnarray}
 where  $e(\phi)$ and
$e(\gamma)$ are the $\phi$ meson and $\gamma$ quantum polarization
four-vectors, $G_R(m)$ is the invariant amplitude free from
kinematical singularities. Since there are no charge particles or
particles with magnetic moments in the process, there is no pole
in $G_R(m)$. Consequently, the function
\begin{eqnarray}
g_R(m)= - 2(pk)G_R(m) = - 2\omega (m) m_\phi G_R(m) \label{gRm}
\end{eqnarray}
is proportional to $\omega (m)$ (at least!) in the soft photon
region.

To describe the experimental spectra similar to the ones
 in
Figs. \ref{myfiga0} and \ref{figf0} \footnote{Note that
$S_R(m)=\Gamma_\phi dBR(\phi\to\gamma R\to\gamma ab\,,\,
m)/dm\,.$},
 the function $|g_R(m)|^2$ should be smooth (almost constant) in the
range $m\leq 0.99$ GeV, see Eq. (\ref{spectrumR}). Stopping the
function $(\omega (m))^2$ at $\omega (990\,\mbox{MeV})=29$ MeV
with the help of the form-factor $1/\left [1+(R\omega (m))^2\right
]$ requires $R\approx 100$ GeV$^{-1}$. It seems to be incredible
to explain the formation of such a huge radius in hadron physics.
 Based on the
large, by hadron physics standard, $R\approx10$ GeV$^{-1}$, one
can obtain an effective maximum of the mass spectrum under
discussion only near 900 MeV. To exemplify this trouble let us
consider the contribution of the isolated R resonance:
$g_R(m)=-2\omega (m) m_\phi G_R\left (m_R\right )$. Let also the
mass and the width of the R resonance equal 980 MeV and 60 MeV,
then
$S_R(920\,\mbox{MeV}):S_R(950\,\mbox{MeV}):S_R(970\,\mbox{MeV}):S_R(980\,\mbox{MeV})
=3:2.7:1.8:1$.

So stopping the $g_R(m)$ function is the crucial point in
understanding  the mechanism  of the production of  $a_0(980)$ and
$f_0(980)$ resonances in the $\phi$ radiative decays.

The $K^+K^-$ loop model $\phi\to K^+K^-\to\gamma R$
\cite{achasov-89} solves this problem in the elegant way: the fine
threshold phenomenon is discovered, see Fig. \ref{g}, where the
universal in $K^+K^-$ loop model function $|g(m)|^2=\left
|g_R(m)/g_{RK^+K^-}\right |^2$ is shown
 \footnote{The forms of
$g_R(m)$ and $g(m)=g_R(m)/g_{RK^+K^-}$ everywhere over the $m$
region are in Refs. \cite{achasov-89} and
\cite{kiselev,achasov-01a} respectively. Emphasize that fitting in
the captions to Figs. \ref{myfiga0} and \ref{figf0} relates only
to the parameters of the scalar mesons ( to  masses, coupling
constants and relative phases of amplitudes of production), not to
$g(m)$, see Refs. \cite{achasov-01,kiselev}. }.

\begin{figure}
\centerline{\epsfxsize=12cm \epsfysize=8cm \epsfbox{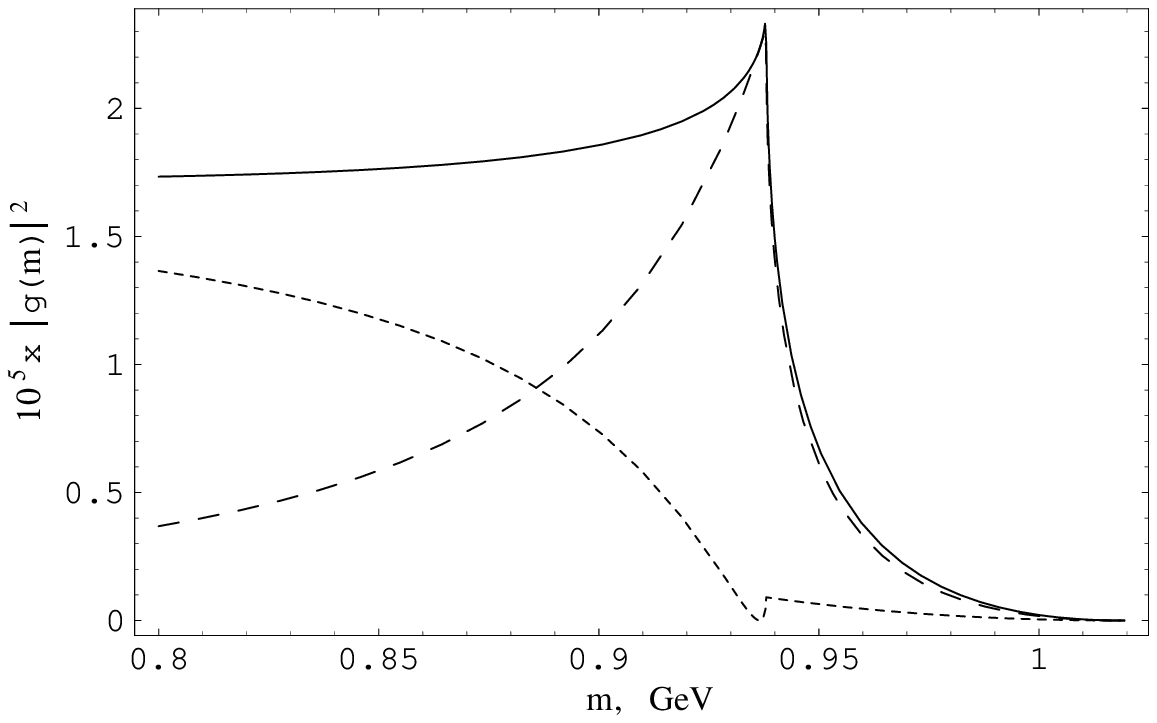}}
 \caption{ The function $|g(m)|^2$ for  $m_{K^+}=469$ MeV is drawn
  with the solid line. The contribution of the imaginary part is drawn with the dashed line.
  The contribution of the real part is drawn with the dotted line.} \label{gg}
\end{figure}

\begin{figure}
\centerline{\epsfxsize=12cm \epsfysize=8cm \epsfbox{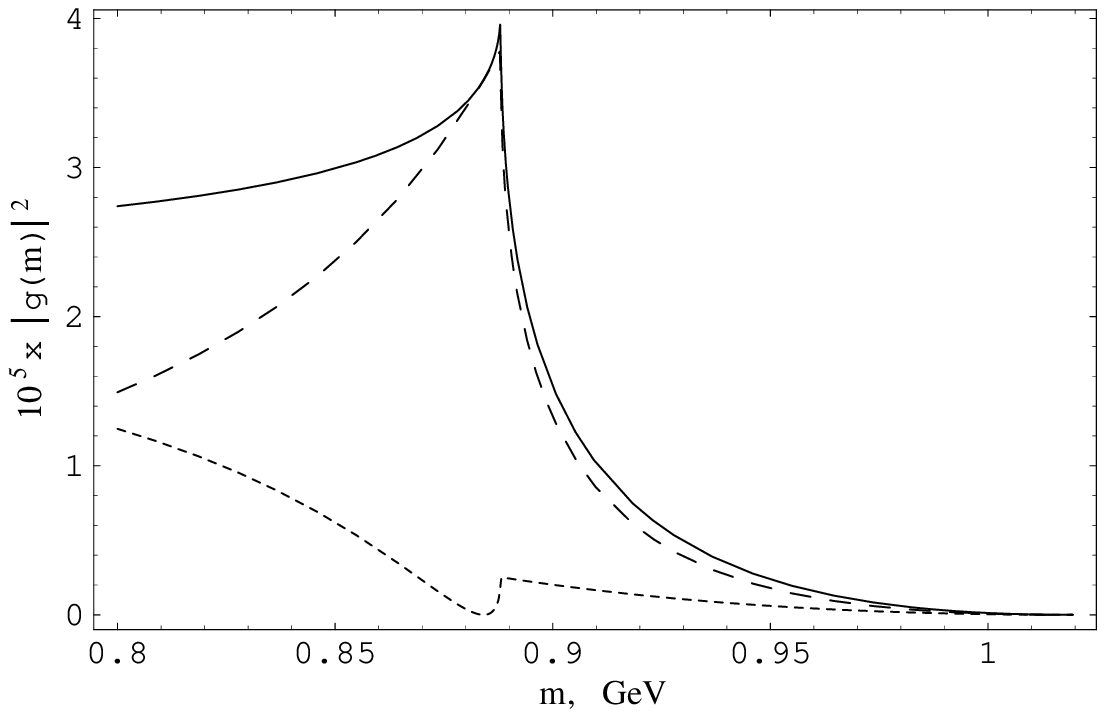}}
 \caption{ The function $|g(m)|^2$ for  $m_{K^+}=444$ MeV is drawn
  with the solid line. The contribution of the imaginary part is drawn with the dashed line.
  The contribution of the real part is drawn with the dotted line.} \label{ggg}
\end{figure}

 To demonstrate the threshold character of this effect we present
 Fig. \ref{gg} and Fig. \ref{ggg} in which the function $|g(m)|^2$
is shown in the case of $K^+$ meson mass is 25 MeV and 50 MeV less
than in reality.

One can see from Figs. \ref{gg} and \ref{ggg} that the function
$|g(m)|^2$ is suppressed by the $(\omega (m))^2$ law in the region
950-1020 MeV and 900-1020 Mev respectively

In the mass spectrum this suppression is increased by one more
power of $\omega (m)$, see Eq. (\ref{spectrumR}), so that we
cannot see the resonance in the region 980-995 MeV \footnote{The
actual absence of a background at a soft photon energy region
$\omega (m)< 112$ MeV ($m > 900$ MeV) owes to gauge invariance
also.}.
 The maximum in the spectrum is effectively shifted to the
region 935-950 MeV and 880-900 MeV respectively.

In truth this means that $a_0(980)$ and $f_0(980)$ resonances are
seen in the radiative decays of $\phi$ meson owing to the $K^+K^-$
intermediate state, otherwise the maxima in the spectra would be
shifted to 900 MeV.

So the mechanism of production of $a_0(980)$ and $f_0(980)$
mesons in the $\phi$ radiative decays is established.

\section{The large \boldmath{$N_C$} expansion of the
\lowercase{\boldmath{$\phi\to\gamma a_0$}} and
\lowercase{\boldmath{$\phi\to\gamma f_0$}} amplitudes}

Both real and imaginary parts of the $\phi\to\gamma R$
 amplitude are caused by the $K^+K^-$ intermediate state. The
imaginary part is caused by the real $K^+K^-$ intermediate state
while the real part is caused by the virtual compact $K^+K^-$
intermediate state, i.e., we are dealing here with the four-quark
transition
\footnote{It will be recalled that the imaginary part
of every hadronic amplitude describes a multi-quark transition.} .
Needless to say, radiative four-quark transitions can happen
between two $q\bar q$ states as well as between $q\bar q$ and
$q^2\bar q^2$ states but their intensities depend strongly on a
type of the transitions. A radiative four-quark transition between
two $q\bar q$ states requires creation and annihilation of an
additional $q\bar q$ pair, i.e., such a transition is forbidden
according to the Okuba-Zweig-Izuka (OZI) rule, while a radiative
four-quark transition between $q\bar q$ and $q^2\bar q^2$ states
requires only creation of an additional $q\bar q$ pair, i.e., such
a transition is allowed according to the OZI rule.

Let us consider this problem from the large $N_C$ expansion
standpoint, using the G.'t Hooft rules \cite{hooft}: $g^2_sN_C\to
const$ at $N_C\to\infty$ and a gluon is equivalent to a
quark-antiquark pair ( $A^i_j\sim q^i\bar q_j$ ). The point is
that the large $N_C$ expansion is the most clear heuristic
understanding of the OZI rule because the OZI forbidden branching
ratio as a rule is suppressed by the factor $N_C^2=9$, but not
$N_C=3$, in comparison with the OZI allowed decay. In our case the
results analysis are even more interesting.

\begin{figure}
\centerline{\epsfxsize=16cm \epsfysize=20cm
\epsfbox{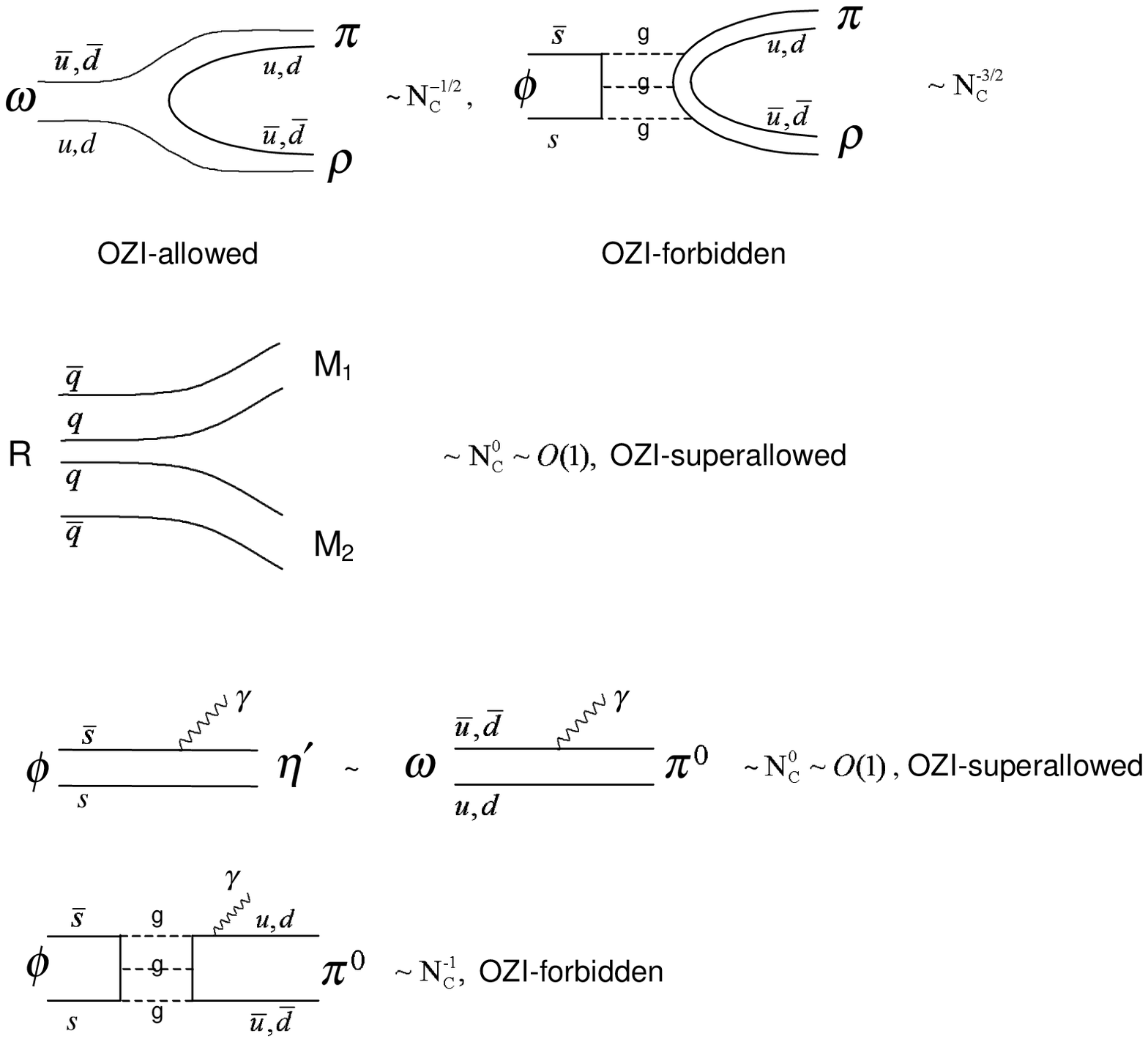}}
 \caption{The large $N_C$ expansion of
some well-known decay amplitudes.} \label{reminder}
\end{figure}

\begin{figure}
\centerline{\epsfxsize=16cm \epsfysize=10cm \epsfbox{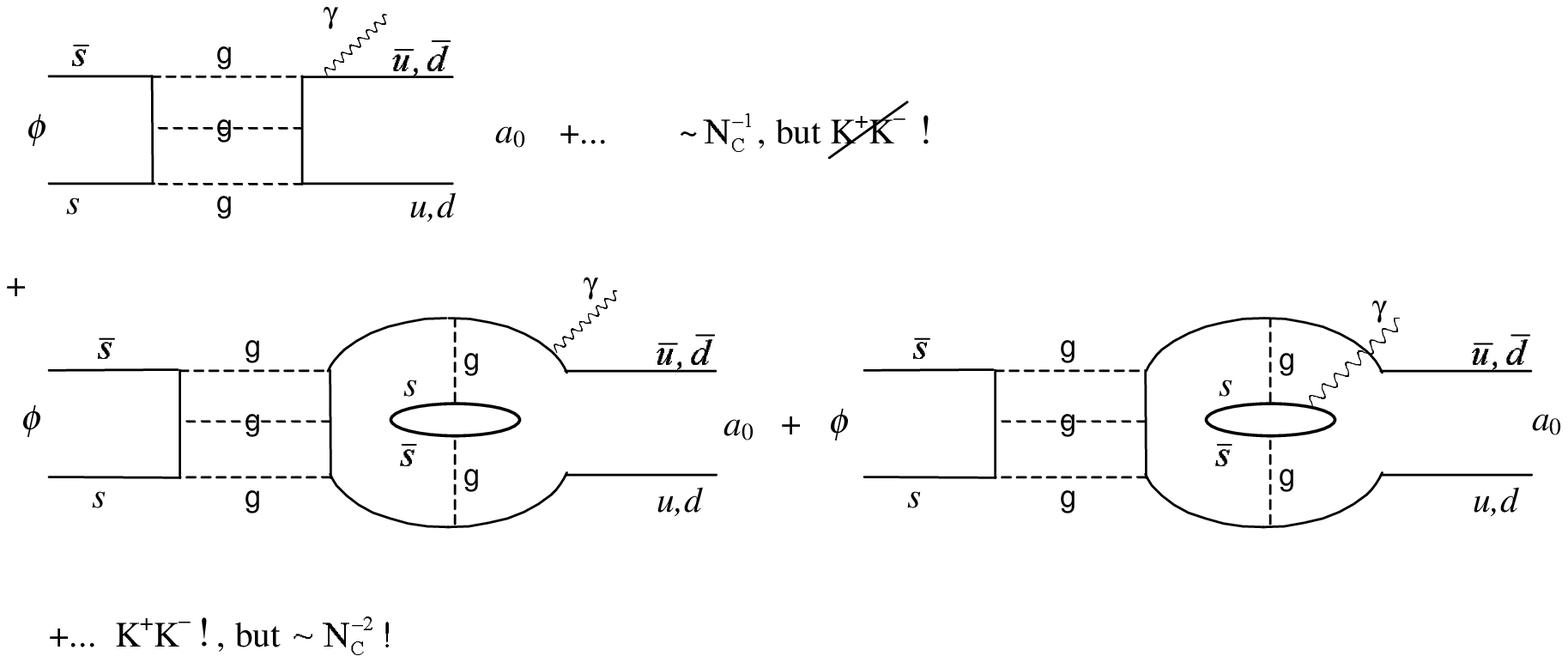}}
 \caption{The large $N_C$ expansion of the $\phi\to\gamma
a_0(980)$ amplitude in the two-quark model $a^0_0(980)=(u\bar u -
d\bar d)/\sqrt{2}$.} \label{a0twonc}
\end{figure}

\begin{figure}
\centerline{\epsfxsize=16cm \epsfysize=20cm
\epsfbox{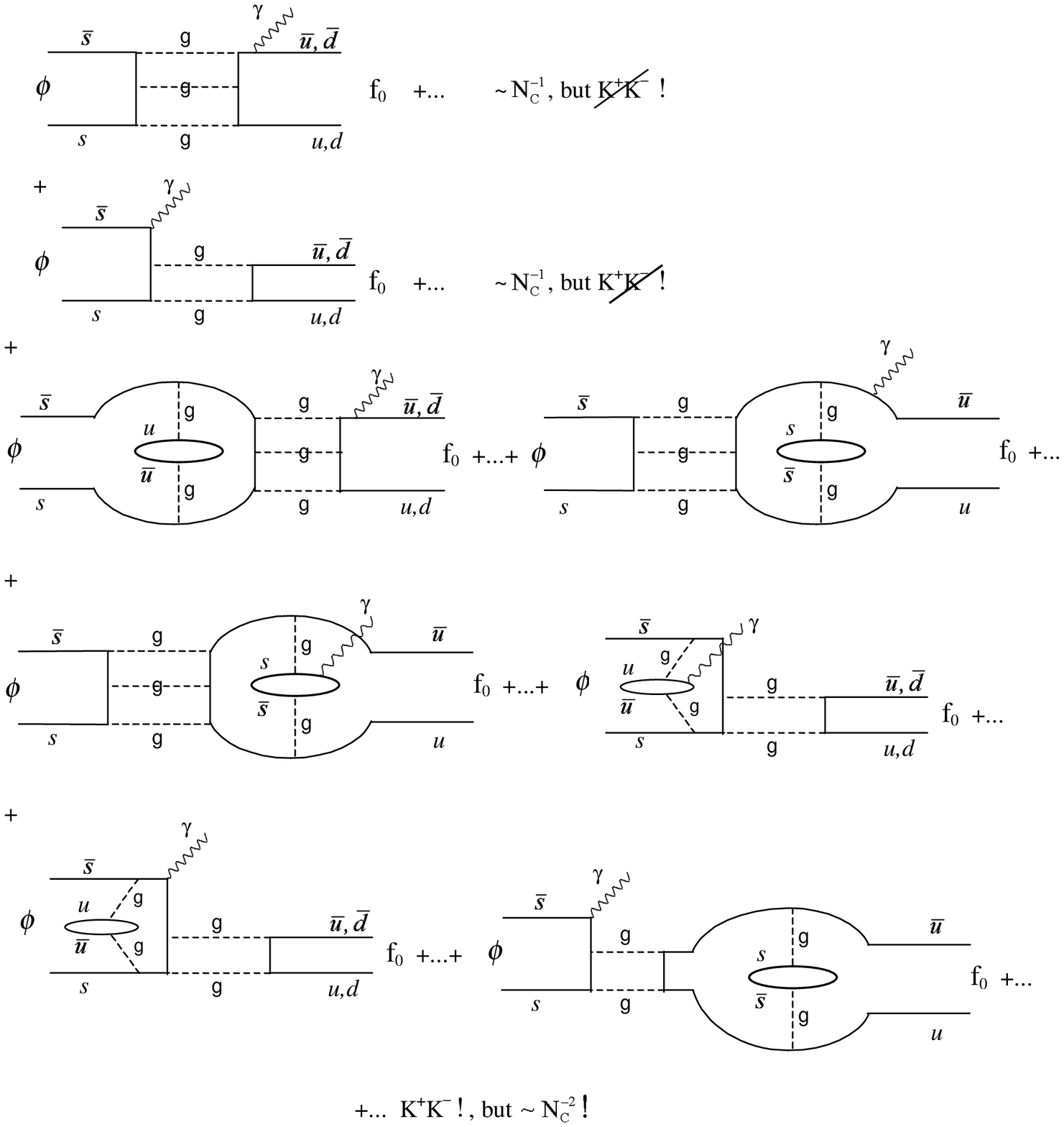}} \caption{The large $N_C$
expansion of the $\phi\to\gamma f_0(980)$ amplitude in the
two-quark model $f_0(980)=(u\bar u + d\bar d)/\sqrt{2}$.}
\label{f0twononstrangenc}
\end{figure}

\begin{figure}
\centerline{\epsfxsize=16cm \epsfysize=10cm
\epsfbox{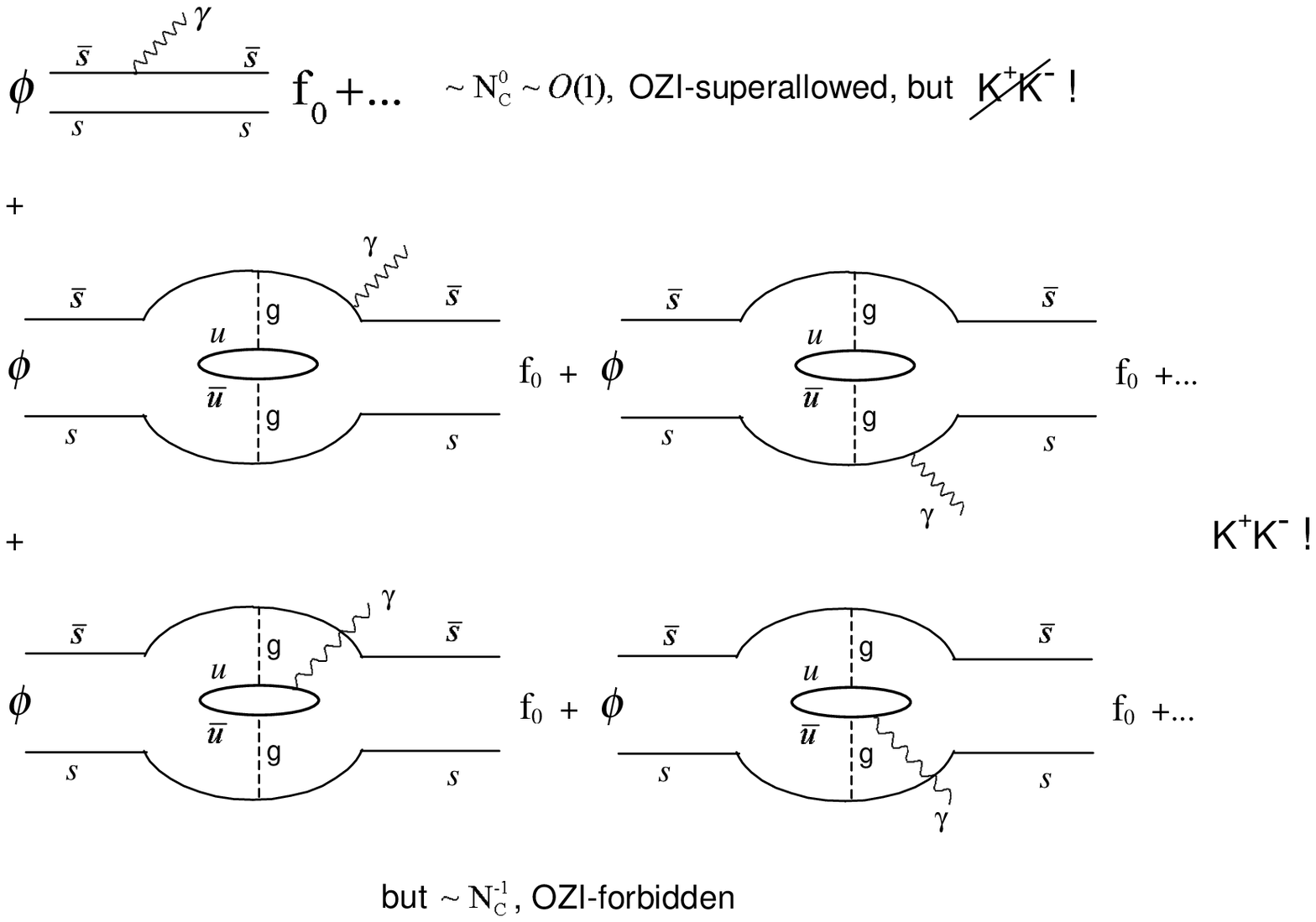}}
 \caption{The large $N_C$ expansion of the $\phi\to\gamma
f_0(980)$ amplitude in the two-quark model $f_0(980)=s\bar s$.}
\label{f0twostrangenc}
\end{figure}

\begin{figure}
\centerline{\epsfxsize=16cm \epsfysize=10cm
\epsfbox{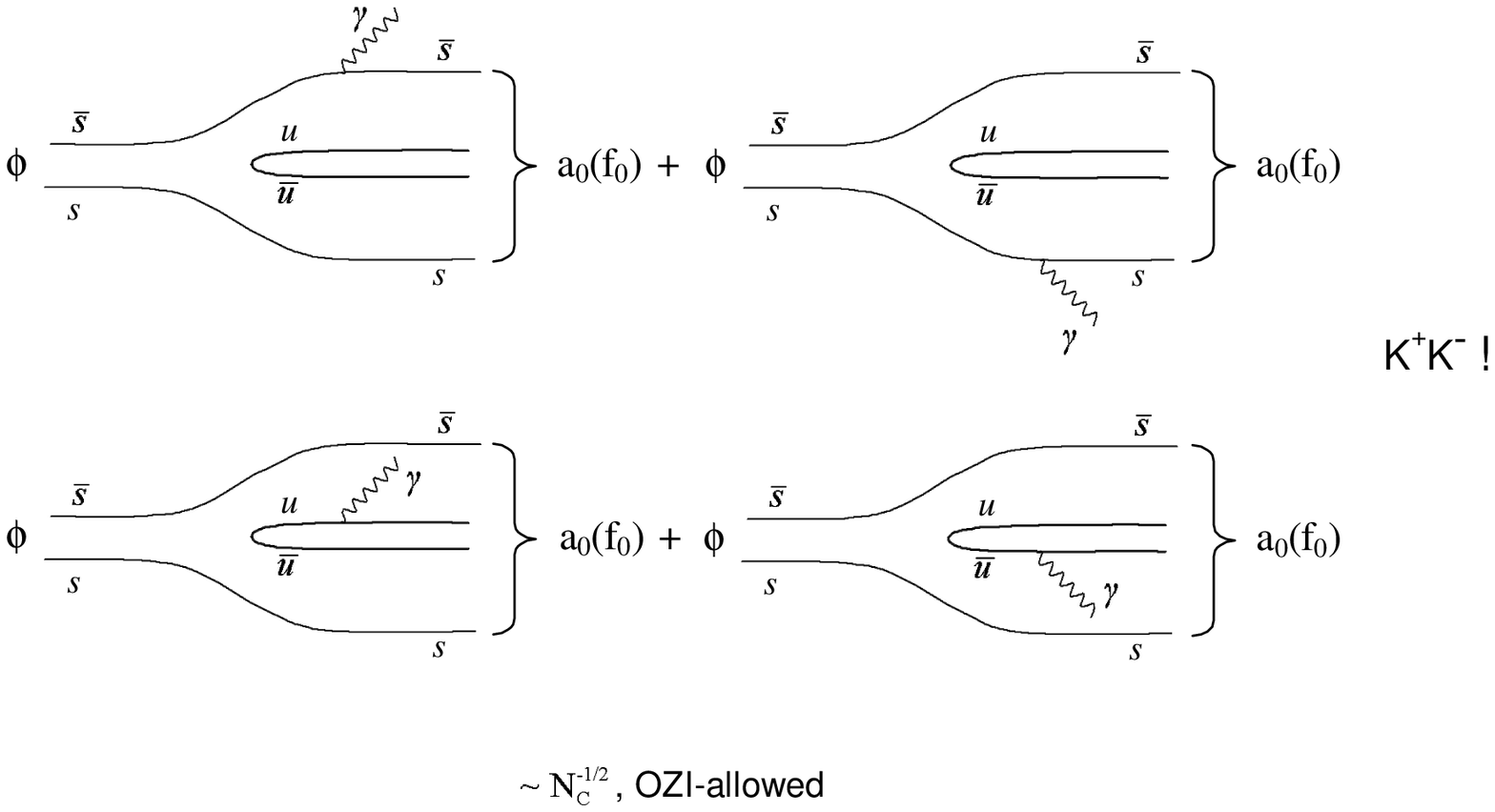}}
 \caption{The large $N_C$ expansion of the $\phi\to\gamma
a_0(980)$ and $\phi\to\gamma f_0(980)$ amplitudes in the
four-quark model $a^0_0(980 =(u\bar ss\bar u - d\bar ss\bar
d)/\sqrt{2}$ and $f_0(980 = (u\bar ss\bar u + d\bar ss\bar
d)/\sqrt{2}$.} \label{a0f0fournc}
\end{figure}

Fig. \ref{reminder} reminds us  of the large $N_C$ expansion of
some well-known decay amplitudes
\footnote{In Figs. \ref{reminder}
- \ref{a0f0fournc} are understood the graphs with the obvious
permutations of the gamma quantum. In addition, in Figs.
\ref{reminder} - \ref{a0f0fournc} are not shown exchanges by
planar gluons which do not change the large $N_C$ behaviour. }

Let us begin our consideration with the $q\bar q$ model.

In the two-quark model $a^0_0(980)=(u\bar u - d\bar d)/\sqrt{2}$,
the large $N_C$ expansion of the $\phi\approx s\bar s\to\gamma
a_0(980)$ amplitude starts with the OZI forbidden transition of
the $1/N_C$ order:  $s\bar s$ annihilation and $u\bar u,\, d\bar
d$ creation, see Fig. \ref{a0twonc}. But its weight is bound to be
small, because this term does not contain the $K^+K^-$
intermediate state, which emerges only in the next to leading term
of the $\left (1/N_C\right )^2$ (!) order for creation and
annihilation of additional $q\bar q$ pairs, see Fig.
\ref{a0twonc}. Note that $\phi\approx s\bar s\to\gamma s\bar
s\to\gamma\eta ' (958)$ transition ( as intensive experimentally
as $\phi\to\gamma a_0(980)$ ) does not require creation of an
additional $q\bar q$ pair at all (the OZI superallowed transition)
and has the $\left (N_C\right )^0$ order, see Fig. \ref{reminder}.

In the two-quark model $f_0(980)=(u\bar u + d\bar d)/\sqrt{2}$,
which involves the $a_0$\,-$f_0$ mass degeneration, the large
$N_C$ expansion of the $\phi\approx s\bar s\to\gamma f_0(980)$
amplitude starts also with the OZI forbidden transition of the
$1/N_C$ order: $s\bar s$ annihilation and $u\bar u,\, d\bar d$
creation, see Fig. \ref{f0twononstrangenc}, whose weight also is
bound to be small, because this term does not contain the $K^+K^-$
intermediate state, which emerges only in the next to leading term
of the $\left (1/N_C\right )^2$ order, see Fig.
\ref{f0twononstrangenc}.

In the two-quark model $f_0(980)\approx s\bar s$, which has the
serious trouble with the $a_0$\,-$f_0$ mass degeneration,  the
$\left (N_C\right )^0$ order transition without creation of an
additional $q\bar q$ pair $\phi\approx s\bar s\to\gamma s\bar
s\to\gamma f_0(980)$
 \footnote{In this regard the $\left
(N_C\right )^0$ order mechanism is similar to the principal
mechanism of the $\phi\approx s\bar s\to\gamma s\bar
s\to\gamma\eta ' (958)$ decay, see Fig. \ref{reminder}.}, see Fig.
\ref{f0twostrangenc},
 is bound to have a small weight in the
large $N_C$ expansion of the $\phi\approx s\bar s\to\gamma
f_0(980)$ amplitude, because this term does not contain the
$K^+K^-$ intermediate state, which emerges only in the next to
leading term of the $1/N_C$ order, i.e., in the OZI forbidden
transition, see Fig. \ref{f0twostrangenc}. Emphasize that the
mechanism without creation  and annihilation of the additional
$u\bar u$ pair cannot explain the $S_{f_0}(m)$ spectrum because it
does not contain the $K^+K^-$ intermediate state!

But if $a^0_0(980)$ and $f_0(980)$ mesons are compact $K\bar K$
states $a^0_0(980)=(u\bar ss\bar u - d\bar ss\bar d)/\sqrt{2}$ and
$f_0(980)= (u\bar ss\bar u + d\bar ss\bar d)/\sqrt{2}$
 respectively, i.e., four-quark states similar     to states of
the MIT-bag model
 \footnote{In the case of the $K\bar K$ bound
states with the binding energy close to 20 MeV , i.e.,in the
extended molecule case, the contribution of the virtual
intermediate $K^+K^-$ states in the $K^+K^-$ loop is suppressed by
the momentum distribution in the molecule, and the real part of
the loop amplitudes are negligible \cite{achasov-97a}. It leads to
the branching ratios \cite{achasov-97a} much less than the
experimental ones . In addition, the $S_R(m)$ spectra in the
$K\bar K$ molecule case are much narrower than the experimental
ones, see the behavior of the imaginary part contribution in Fig.
\ref{g}.},
 the large $N_C$
expansions of the $\phi\approx s\bar s\to\gamma a_0(980)$ and
$\phi\approx s\bar s\to\gamma f_0(980)$ amplitudes start  with the
OZI allowed transitions of the $\left (1/N_C\right )^{-1/2}$
order, which require only creation the additional $u\bar u$ pair
for the $K^+K^-$ intermediate state, see Fig. \ref{a0f0fournc}.

Thus, it is shown that the $\phi\to\gamma a_0$ decay intensity in
the two-quark model  and the $\phi\to\gamma f_0$ decay intensity
in the $f_0=(u\bar u + d\bar d)/\sqrt 2$ model are suppressed by
the factor $1/N_c^3$ in comparison with the ones in the four-quark
model, but in the $f_0=s\bar s$ model the $\phi\to\gamma f_0$
decay intensity is suppressed only by the factor $1/N_c$ in
comparison with the one in the four-quark model. It will be
recalled that a OZI allowed hadronic decay amplitude, for example,
the $\rho\to\pi\pi$ amplitude, see Fig. \ref{reminder}, has the
$\left (1/ N_C\right ) ^{-1/2}$ order, that is, a OZI allowed
hadronic decay intensity is suppressed by the factor $1/N_c$ in
comparison with a OZI superallowed one.

\section{Conclusion}

In summary the fine threshold phenomenon is discovered, which is
to say that the $K^+K^-$ loop mechanism of
 the $a_0(980)$ and $f_0(980)$ scalar meson production in the
$\phi$ radiative decays is established at a physical level of
proof. The case is rarest in hadron physics. This production
mechanism is the four-quark transition what constrains the large
$N_C$ expansion of the $\phi\to\gamma a_0(980)$ and $\phi\to\gamma
f_0(980)$ amplitudes and gives the new strong (if not crucial)
evidences in favor of the four-quark nature of $a_0(980)$ and
$f_0(980)$ mesons.

\section*{Acknowledgement}

I thank A.V. Kiselev very much for the help.

A preliminary version of this paper has been reported as Invited
Talk  on February 25, 2003 at
 "International Symposium on Hadron Spectroscopy, Chiral Symmetry and
 Relativistic Description of Bound Systems",
February 24 - 26, 2003, Tokyo, Japan.
 I would like to thank  Professor Ishida-San, Professor
Takamatsu-San, Professor Tsai-san and Professor Tsuru-San very
much for the Invitation, the financial support and the generous
hospitality. I would like to thank all participants of
"International Symposium on Hadron Spectroscopy, Chiral Symmetry
and  Relativistic Description of Bound Systems", February 24 - 26,
2003, Tokyo, Japan, and "KEK Workshop on hadron
 spectroscopy and chiral particle search in $J/\psi$ decay data at BES",
February 28 -  March 1, 2003, Tsukuba, Japan, for a creative
atmosphere, which promoted writing the final version of the paper.

I would like also to thank Referee in Nuclear Physics A, whose
questions promoted the emergence of many elucidating and critical
comments.

This work was supported in part by RFBR, Grant No 02-02-16061.

\newpage

\end{document}